\documentclass{article}

\usepackage{amsmath}
\usepackage{comment}

\usepackage[preprint]{neurips_2026}

\usepackage[utf8]{inputenc} 
\usepackage[T1]{fontenc}    
\usepackage{hyperref}       
\usepackage{url}            
\usepackage{booktabs}       
\usepackage{amsfonts}       
\usepackage{nicefrac}       
\usepackage{microtype}      
\usepackage{xcolor}         
\usepackage{enumitem}
\usepackage{graphicx}
\usepackage{multirow}
\usepackage{tabularx}

\RequirePackage{listings}
\definecolor{dkgreen}{rgb}{0,0.6,0}
\definecolor{gray}{rgb}{0.5,0.5,0.5}
\definecolor{mauve}{rgb}{0.58,0,0.82}

\title{Machine Learning Interatomic Potentials: Advancing Open-Source Software for Efficient and Scalable Molecular Simulation}

\author{
  Christoph Brunken \And
  Titouan Cormier \And
  Lucien Walewski \And
  Marco Carobene \And
  Yessine Khanfir \And
  Zachary Weller-Davies \And
  Miguel Bragança \And
  Armand Picard \And
  Adrien Pichard \And
  Leon Wehrhan \And
  Heloise Chomet \And 
  Eszter Varga-Umbrich \And
  Marie Bluntzer \And
  Massimo Bortone \And
  Valentin Heyraud \And
  Silvia Acosta-Guti\'{e}rrez \And
  Jules Tilly \And
  Olivier Peltre \AND
  {\normalfont InstaDeep} \\ \\
  1 Triton Square, London, NW1 3BF, United Kingdom \\
  \texttt{\{c.brunken,j.tilly,o.peltre\}@instadeep.com} \\
}

\begin{document}

\maketitle

\begin{abstract}
Machine learning interatomic potentials (MLIPs) enable atomistic simulations with near \textit{ab initio} accuracy at significantly reduced computational cost, but their broader adoption is often limited by fragmented tooling, limited scalability, and inflexible software design.
We present \textit{mlip} v2, a new generation of the \textit{mlip} library that advances efficient and scalable molecular simulation through a unified and extensible framework. The new release features a targeted API redesign with improved modularity and control, enabling flexible customization of training, data processing, and simulation workflows. It further integrates a new high-performance backend for equivariant operations, {\it e3j}, significantly accelerating model inference and simulations.
In addition, the framework introduces a range of entirely new capabilities, including the eSEN architecture with a Mixture-of-Experts formulation for scalable training on large and diverse datasets, improved handling of electrostatics through more physically grounded charge modeling and long-range interaction treatment, and advanced simulation features such as NPT ensembles and nudged elastic band methods. Together, these extensions significantly broaden the scope of MLIP applications, enabling efficient modeling of complex, reactive, and out-of-equilibrium systems, and bridging the gap between ML research and practical molecular simulation applications. The library is available on \href{https://github.com/instadeepai/mlip}{GitHub} and on \href{https://pypi.org/project/mlip/}{PyPI} under the Apache license 2.0.
\end{abstract}

\newpage
\section{Introduction}

Machine learning interatomic potentials (MLIPs) have emerged as a powerful approach for atomistic simulation, enabling near \textit{ab initio} accuracy at significantly reduced computational cost. By learning effective energy and force representations from quantum mechanical data, e.g., density functional theory (DFT), MLIPs provide a practical alternative to traditional force fields while remaining orders of magnitude faster than electronic structure methods. However, the adoption and development of MLIPs in practice are often hindered by the lack of unified, high-quality software infrastructure that supports (i) efficient simulation, (ii) ease of use for applications, and (iii) rapid methodological innovation.

To address this gap, the {\it mlip} library~\cite{brunken2025machinelearninginteratomicpotentials} was introduced in May 2025, a unified framework for training and deploying MLIP models, e.g., in molecular dynamics (MD) simulations. The library is entirely based on JAX, benefiting from just-in-time XLA (Accelerated Linear Algebra) compilation,
and the efficient integration with the JAX-MD simulation backend~\cite{schoenholz2020jaxmdframeworkdifferentiable} allowing for state-of-the-art molecular dynamics (MD) simulation speeds. The \textit{mlip} library has since seen broad adoption, with thousands of downloads and usage in a growing number of downstream research projects (for example, see Ref.~\cite{priyadarshini2025diversitydriven, Cui2026.03.18.712561, huang2025skillpuzzler}). This first version of the library demonstrated the feasibility of a unified software stack that connects model training, inference, and molecular simulation in an efficient and easy-to-adopt framework to enable rapid prototyping and testing of novel ideas in the field of MLIPs. Moreover, it supports data-parallel training (even on a multi-host setup) on GPUs and TPUs, as well as batched inference, batched MD simulations, and batched energy minimizations. The library was also integrated into MLIPAudit, a JAX-compatible benchmarking toolset for MLIPs~\cite{wehrhan2025mlipauditbenchmarkingtoolmachine}.

While version 1 (v1) provides a strong foundation for MLIP development, it was designed as a first-generation system, and naturally left room for further expansion.
In this work, we present \textit{mlip} v2, a new generation of the \textit{mlip} library that addresses limitations through a targeted redesign, while preserving familiar interfaces and usage patterns. The redesigned architecture improves composability and control across end-to-end MLIP pipelines, including data processing, training, and molecular simulation. It also integrates {\it e3j}, a fully open-source high-performance backend for equivariant operations, with dedicated CUDA and Pallas kernels enabling significantly faster inference and MD simulations on both GPUs and TPUs (due for full release in June 2026).

Beyond performance and usability improvements, \textit{mlip} v2 significantly expands the scope of supported methodologies and scientific applications:

\begin{enumerate}[nosep]
    \item \textbf{Model architectures:} In addition to the models already included in v1, we integrate the eSEN~\cite{passaro2023escn, fu2025learningsmoothexpressiveinteratomic, wood2026umafamilyuniversalmodels} architecture with an optional Mixture-of-Experts formulation, enabling scalable training on large and diverse datasets while preserving efficient inference. Most methods, embeddings, activations and layers are now interchangeable between model architectures, maximizing modularity for novel method development. 
    \item \textbf{Scientific capabilities:} The framework further extends its capabilities with support for advanced simulation and training paradigms, including NPT ensemble simulations (isothermal-isobaric), Hessian-label model training, and nudged elastic band (NEB) methods for transition state search.
    \item  \textbf{Charge predictions and electrostatics:} All models now support atomic partial charge prediction, and the treatment of electrostatics is significantly improved through explicit charge conditioning and long-range interaction modeling, enabling more physically faithful representations of charged systems.
    \item \textbf{Fine-tuning:} Finally, multi-head fine-tuning is generalized across architectures, improving flexibility and consistency when adapting pretrained models to downstream tasks.
\end{enumerate}

Overall, \textit{mlip} v2 substantially expands the capabilities and performance of the \textit{mlip} framework while introducing a more modular and flexible architecture for future research and development. The API has been redesigned to improve composability and extensibility while preserving a familiar interface for core workflows and minimizing breaking changes. To support adoption, we also provide a comprehensive migration guide from v1. Together, these improvements establish \textit{mlip} v2 as a scalable and adaptable foundation for ML-based molecular simulation.

\section{Background and related work}

MLIPs have rapidly placed themselves among the most promising AI based avenues for enhancing molecular discovery, spanning across materials discovery, bio-molecular sciences, and chemical engineering. In this library, we have so far focused on families of models based on Graph Neural Networks (GNNs) that preserve physical symmetries -- at the very least invariance to rotation and translations in energy predictions, but oftentimes also equivariance of spatial output and latent information throughout the network. While equivariance usually comes at a computational cost, it has also been shown to improve data efficiency and stability in simulations~\cite{Batzner2022, brehmer2024doesequivariancematterscale}.

A first example of symmetry-preserving methods includes distance-based or invariant GNNs. These methods rely on interatomic distances to learn rotationally invariant energy predictions (e.g., SchNet~\cite{Schtt2018, schutt2019schnetpack, schutt2023schnetpack}, AIMNet2~\cite{Anstine2025}, Crystal Graph Convolutional Neural Network~\cite{Xie2018}). To achieve equivariance, one needs to preserve geometric information throughout the network, for instance, by equivariantly updating edge features between each message passing layers~\cite{satorras2022enequivariantgraphneural} or guaranteeing inter-layer equivariance through computation of angular features (e.g., DimeNet~\cite{gasteiger_dimenet_2020, gasteiger_dimenetpp_2020}, GemNet~\cite{GemNet, gasteiger2024gemnetuniversaldirectionalgraph}, ViSNet~\cite{Wang2024visnet, Wang2024ai2bmd}). This can also be achieved through general steerable 3D convolutions, generally built on the formalism of Clebsch-Gordan tensor products~\cite{weiler20183dsteerablecnnslearning, kondor2018clebschgordannetsfullyfourier, batatia2022designspacee3equivariantatomcentered} to achieve arbitrary orders of representation of geometric features (e.g., NequIP~\cite{Batzner2022}, the Tensor Field Network~\cite{thomas2018tensorfieldnetworksrotation}, MACE~\cite{batatia2023macehigherorderequivariant, kovács2025maceofftransferableshortrange, batatia2024foundationmodelatomisticmaterials, Kovcs2023, batatia2026macepolar1polarisableelectrostaticfoundation}). Some approaches have also proposed a modified version of the tensor product to achieve higher computational efficiency, such as projections to a 2D domain for faster convolutions, such as eSEN~\cite{passaro2023escn, luo2024enablingefficientequivariantoperations, fu2025learningsmoothexpressiveinteratomic, wood2026umafamilyuniversalmodels}, or by rewriting the tensor product as integrals of the feature vectors' corresponding signal on a sphere~\cite{xie2025the, xie2026asymptoticallyfastclebschgordantensor, heyraud2026integralformulasvectorspherical}. Finally, multiple transformer based methods have also been developed to strictly preserve equivariance (see, e.g., Ref.~\cite{liao2023equiformer, liao2024equiformerv, li2026eformer, huang2026e2formerv2ontheflyequivariantattention}), or to learn it based on data augmentation or targeted loss constraints (see, e.g., Ref.~\cite{qu2026recipescalableattentionbasedmlips, elhag2026learninginteratomicpotentialsexplicit}).

\section{Library overview and updates}

\subsection{Design philosophy}

In its first version, the \textit{mlip} library was designed as an end-to-end toolbox for MLIP models, covering the full workflow from data processing and model development to training and deployment in inference tasks such as MD simulations. The design was guided by three core principles. First, \textbf{ease of use} was prioritized to lower the barrier to entry for application-focused non-expert users, e.g., by providing sensible default configurations and extensive tutorials. Second, \textbf{extensibility} was enabled through a modular architecture that decouples the core building blocks of an end-to-end MLIP pipeline, connected via minimal, clearly defined interfaces. Third, \textbf{inference efficiency} was treated as a key priority, as we identified it as an essential factor for pushing MLIP models towards relevant industrial applications, in particular, in computational biology. This is realized through the tight integration of our JAX-based models with the JAX-MD simulation backend, along with native support for batched simulations, energy minimization, and inference workloads.

With continued development and broader usage of the library, we identified several opportunities to further improve the original design, particularly with respect to extensibility.
While the modular structure of \textit{mlip} v1 enabled a range of customization options, certain components, such as model call signatures, proved less flexible than desired for emerging advanced use cases and the new features introduced in this work. 
Similarly, the data processing pipeline was primarily optimized for standard training workflows, and extending it to more complex scenarios such as inference-specific preprocessing, multi-dataset training, or multi-head fine-tuning required additional effort.
In addition, many model implementations were initially adapted from existing open-source projects, which allowed for rapid development but did not fully exploit the potential for reusable building blocks and a consistent architectural pattern.
These observations motivated a set of targeted API and implementation refactors in \textit{mlip} v2, aimed at improving flexibility and making the library a more powerful and convenient toolbox for both applied users and researchers developing new or customized MLIP models.

On the performance side, our focus on inference efficiency led to the identification of new requirements, in particular the need for a custom low-level backend to further optimize execution times. This motivated the development and integration of {\it e3j} into the library.

At the same time, we place a strong emphasis on maintaining a stable and familiar user experience. Despite internal changes, the overall structure and feel of the API remain largely consistent, with only minimal adjustments for existing users. In this sense, v2 is designed to make the library more powerful and flexible for advanced users, while preserving the simplicity and usability that applied researchers rely on, and also delivering tangible benefits for them in the form of significant speed improvements and additional features.

\subsection{Usage and migration}

\textit{mlip} v2 has been released as version 0.2.0 of the library on GitHub~\footnote{\url{https://github.com/instadeepai/mlip}} and PyPI~\footnote{\url{https://pypi.org/project/mlip}} in May 2026, and can be installed easily via \texttt{pip}.

A detailed overview of API and code updates can be found either in our extensive migration guide that is part of our code documentation~\footnote{\url{https://instadeepai.github.io/mlip}} or in Appendix~\ref{app:api_changes}. As mentioned above, for most applied research workflows, migration from v1 to v2 will require minimal effort.

A central component of the redesigned v2 code base is the introduction of a unified \texttt{Graph} class, which replaces the previously used \texttt{jraph.GraphsTuple} objects. This change removes the dependency on \textit{jraph}~\cite{jraph2020github}, a project archived in May 2025, and provides a more maintainable and future-proof foundation. Beyond this, the \texttt{Graph} class offers a more expressive and well-documented representation of graph data, with built-in methods for common operations such as computing edge vectors from stored positions and connectivity. In v2, it serves as the core data structure throughout the library, consistently representing model inputs, outputs, and even optionally intermediate latent features. This unified representation allows us to standardize all major model components around a \texttt{Graph\,$\rightarrow$\,Graph} interface, leading to cleaner composition of model blocks and greater flexibility in designing and reusing architectures. Finally, its object-oriented design makes it straightforward for advanced users to extend via inheritance (e.g., to add new input or output features), while updated data processing pipelines follow similar patterns, allowing customization with minimal code duplication.

Building on these changes, all core model implementations in v2 were refactored with this unified design in mind, adopting consistent interfaces and reusable building blocks. In addition, we integrated the {\it e3j} library to accelerate MACE and NequIP models (see section~\ref{sec:e3j}). For backwards compatibility, the original v1 model implementations are still available within a dedicated module. However, they are now exposed through the updated and slightly extended v2 \texttt{ForceField} interface, ensuring that models trained with the previous codebase remain usable. We note that this support is intended as a transitional solution, and plan to phase out the legacy implementations in future releases as part of ongoing efforts to simplify and modernize the code base.

\section{Accelerating equivariant operations with {\it e3j}} \label{sec:e3j}

In order to optimize runtime on a large span of hardware, the library supports the open-source
{\it e3j} backend. The backend includes dedicated kernels for equivariant operations in both Pallas and CUDA for efficient implementation of MLIP models across both TPUs and GPUs.

\begin{figure}[ht]
    \centering
    \includegraphics[width=\linewidth]{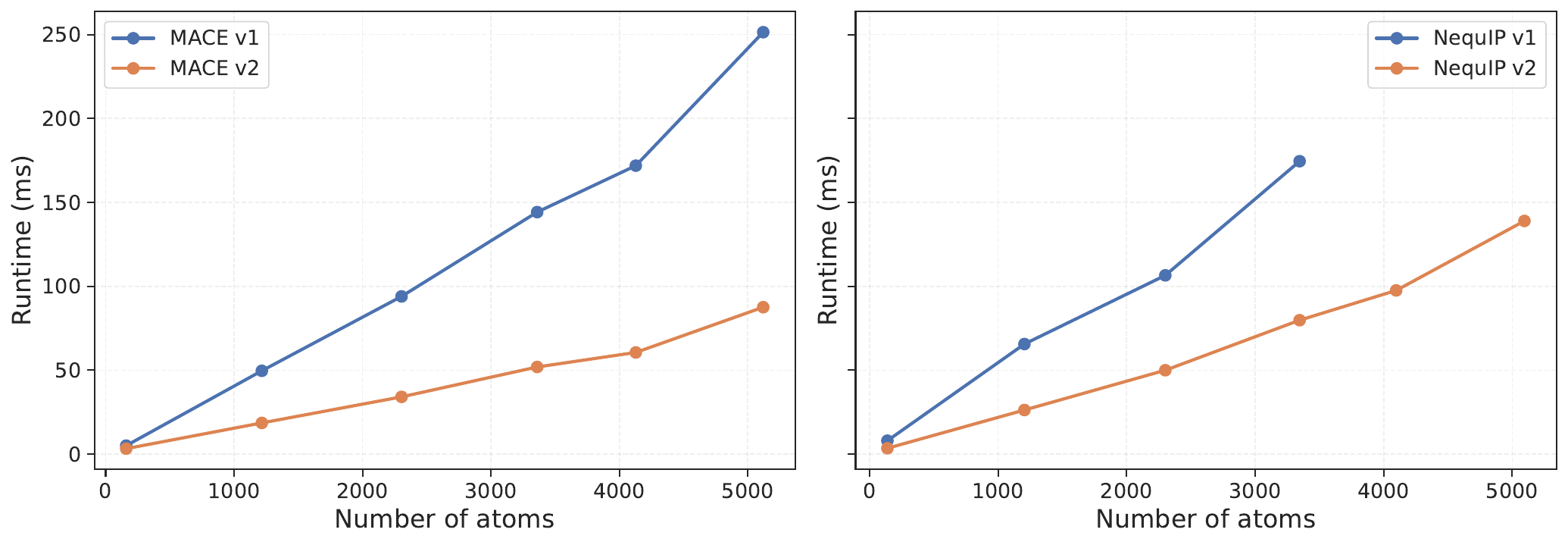}
    \caption{Results of end-to-end runtime benchmarks on MACE and NequIP, comparing the {\it mlip}  v1 and {\it mlip}  v2 versions, showcasing clear speed-ups across the board for both models. For this visualization, we show the Pallas implementation for MACE and the CUDA one for NequIP, as these represent the fastest options at the time of this writing. Further speed-up and scale should be expected following the full release of the {\it e3j} library.}
    \label{fig:e3j}
\end{figure}

This is particularly relevant for MACE and NequIP, which both rely on the Clebsch-Gordan Tensor Product for their internal operations, which usually constitute a significant computational bottleneck. We include with the library examples of training MACE and NequIP with the {\it e3j} backend. These achieve consistent performance boost over the previous version of the library (up to 3x). See Figure~\ref{fig:e3j} for an end-to-end runtime benchmark. It is worth noting that the version of {\it e3j} deployed in the {\it mlip} library is still work in progress. Further work is ongoing to improve fusion of the message passing kernel which should bring additional speed-ups an ability to scale model deployment to larger systems. 

\section{Extended Scientific Capabilities}

\subsection{Treatment of electrostatics and long-range interactions}

We introduce new electrostatics capabilities in \textit{mlip} v2, aimed at improving the modeling of physically relevant interactions. In particular, the library now supports enhanced treatment of long-range electrostatic effects as well as global charge conditioning, enabling more accurate and flexible simulations of charged systems. In the following, we describe these two components in more detail.

Different approaches to modeling long-range interactions in MLIPs have been proposed in the literature~\cite{Unkeetal2019, brunken2024machine, Anstineetal2025, batatia2026macepolar1polarisableelectrostaticfoundation}, where models typically rely on self-predicted atomic partial charges to compute the Coulomb electrostatic contribution $E_{LR}$ to the total energy. Atomic partial charges are obtained via supervised learning, where reference charges (derived from a chosen electronic structure method) are used as targets in the loss function. To ensure physical consistency, the predicted partial charges $q_i$ are constrained such that their sum matches the total system charge $Q_{tot}$, thereby guaranteeing charge conservation,

\begin{equation}
\label{eq:charge_correction}
\tilde{q}_i = q_i - \frac{1}{N_{\mathcal{S}}} \bigg( \sum_{j \in \mathcal{S}} q_j -Q_{tot} \bigg) \quad .
\end{equation}

Here, $\tilde{q}_i$ denotes the corrected partial charge of atom $i$, $q_i$ the raw predicted charge, $\mathcal{S}$ the set of atoms in the system, and $N_{\mathcal{S}}$ its cardinality.
\textit{mlip} v2 currently implements a modified Coulomb interaction term $E_{LR}$ to account for long-range electrostatic interactions, following the formulation introduced in PhysNet~\cite{Unkeetal2019},

\begin{equation}
\label{eq:long_range_interaction}
E_{LR} = \sum_{i,j \in \mathcal{S}} k_e \frac{\tilde{q_i} \tilde{q_j}}{\chi(r_{ij})} \quad ,
\end{equation}

Here, $k_e = \frac{1}{4\pi \epsilon_0}$ is Coulomb’s constant, with $\epsilon_0$ denoting the vacuum permittivity, and $r_{ij}$ the distance between atoms $i$ and $j$.
To avoid divergences of the Coulomb interaction at short interatomic distances, a modified distance function $\chi(r_{ij})$ is used. This surrogate distance matches the true distance $r_{ij}$ beyond a soft-core radius $r_0$, while smoothly regularizing the interaction at shorter distances, as proposed in PhysNet~\cite{Unkeetal2019}.

We emphasize that the new modular API design also makes it straightforward to add custom energy heads, i.e., components that map the rich \texttt{Graph} outputs of the network to the final energy prediction. This facilitates the integration of user-defined long-range interaction models.

Furthermore, we introduce a mechanism for global charge conditioning in \textit{mlip} v2, similar to those developed in UMA~\cite{wood2026umafamilyuniversalmodels}. We observed that MLIP performance can degrade when modeling systems with varying global charge (see Appendix~\ref{app:global_charge}), as differences in formation energies across charge states introduce additional complexity. Incorporating total charge information can therefore improve predictive accuracy in such settings. \textit{mlip} v2 enables this by adding an embedding of the total system charge, which complements the standard atomic number embedding. Both embeddings are concatenated and passed through a lightweight MLP to produce part of the initial model features.

\subsection{Training with Hessian labels} \label{sec:hessian}

Second-order derivatives of the energy with respect to positions are essential
predictions in many downstream tasks, either explicitly 
(e.g., vibrational frequencies or zero point energy) or implicitly via 
running Hessian estimates (e.g., geometry relaxation with BFGS~\cite{BFGS-Fletcher70, BFGS-Goldfarb70, BFGS-Shanno70} or 
transition state search with the dimer method~\cite{Henkelman99}).
Incorporating higher-order derivatives of the energy with respect to atomic coordinates during the training has therefore gained a lot of interest recently, with various 
approaches being developed, either via automatic differentiation~\cite{rodriguez2025does, rodriguez2026projected} or
direct Hessian prediction~\cite{HIP}. 
While forces define the slope of the Potential Energy Surface (PES), 
second-order derivatives carry broader curvature information around each training geometry, further highlighting and differentiating local minima, maxima, saddle points or inflections. This additional information has motivated the use of subsampled Hessian rows during training as a distillation strategy for large foundation models~\cite{amin2501towards}, an approach adopted in \textit{mlip} v2.



Predicting the full Hessian matrix $\mathbf{H}$ for each graph in a batch, and then backpropagating their variations with respect to parameters 
during training would be prohibitively computationally expensive. 
Differentiating all forces $F_i$ with respect to all atomic coordinates $r_i$ scales quadratically with respect to system size,

\begin{equation}
\label{eq:full_hessian}
\mathbf{H}_{ij} = \frac{\partial^2 E}{\partial {\bf r}_{i} \partial {\bf r}_{j}} =
-\frac{\partial {\bf F}_i}{\partial {\bf r}_j} \quad .
\end{equation}

To address this limitation, several works in the literature have explored computing Hessian rows via Vector-Jacobian Products (VJP) and Hessian-Vector Products (HVP), as an alternative to directly learning the full Hessian~\cite{rodriguez2026projected, amin2501towards}.


During training, \textit{mlip} follows the approach of~\cite{amin2501towards} 
by differentiating only selected force components ${\bf F}^{\mathcal{(S)}}$ with respect to all atomic coordinates. For each graph $g$ in the batch, and every row index 
$\rho = 1\dots R$, the subsampling consists of a node index 
$i_{\rho,g}$ chosen within $g$, 
and force predictions on the indices $i_{\rho,g} \in \mathcal{S}$ 
are summed over distinct graphs of the batch.
Reverse-mode automatic differentiation (VJP) on these predictions 
then effectively yields a matrix of size $R \times 3n$, where $n$ is the total 
number of nodes,

\begin{equation}
\label{eq:sampled_hessian2}
\mathbf{H}_{\rho j}^{\mathcal{(S)}} = 
-\frac{\partial {\bf F}_{\rho}^{\mathcal{(S)}}}{\partial {\bf r}_j}
\quad{\rm where}\quad 
{\bf F}_\rho({\bf r}) = \sum_{g\,\in\,{\tt graphs}} {\bf F}_{i_{\rho,g}} \quad .
\end{equation}

Since distinct graphs do not interact, 
each row ${\bf H}_\rho$ is the concatenation of Hessian rows coming from different 
graphs $g$ in the batch, which
is how the Hessian labels are preprocessed in practice.
This sampling-based approximation makes it feasible to include larger and more diverse systems in the training process, thereby broadening the range of accessible reaction types and chemical elements.

The \textit{mlip} v2 library includes the necessary data handling components to efficiently work with Hessian-based training signals in a batched setting. The modularity of its data processing 
pipeline allows to seamlessly hook into both the system pre-processing stage 
(padding full Hessian matrices to a common width) 
and the graph post-processing stage 
(subsample distinct indices at each epoch, 
forward them to the final prediction stage via the \texttt{Graph} object).
Note that in addition to subsampled Hessian training, we also support full Hessian matrix inference. For validation, we present a brief study to assess the correctness of our Hessian label training approach in Appendix~\ref{app:hessian}.

\subsection{Multi-head fine-tuning}

MLIPs pretrained on large and chemically diverse datasets can be specialized to downstream applications through fine-tuning, for example, to a specific chemistry, level of theory or property of interest. A central challenge in this setting is reconciling the often incompatible labeling conventions across datasets, such as different DFT functionals, basis sets or label availability, while avoiding catastrophic forgetting of the original training distribution. Building on multi-head fine-tuning approaches, introduced for MLIPs in~\cite{batatia2023macehigherorderequivariant, Batatia2025}, \textit{mlip} v2 introduces a unified multi-head fine-tuning framework that addresses both points and is agnostic to the model architecture. 

The framework is built on two design choices: a shared equivariant backbone paired with per-dataset readout heads, and a single implementation reused across all model architectures. For a graph $g$ originating from dataset $d_g \in \{1, \ldots, D\}$, the predicted total energy is given by
\begin{equation}
    \label{eq:multihead_readout}
    \tilde{E}_g = \sum_{i\,\in\,g} \left[ \mathcal{R}_{d_g}(h_i) + E_0^{(d_g)}(z_i) \right] 
\end{equation}
where $h_i \in \mathbb{R}^F$ is the latent feature of atom $i$ produced by the shared backbone, $\mathcal{R}_{d}$ is the readout for dataset $d$, and $E_0^{(d)}$ is the dataset-specific table of reference atomic energies indexed by atomic number $z_i$. The same mechanism is reused across all supported architectures (MACE, NequIP, ViSNet, and eSEN), so extending a model with an additional head requires no architecture-specific code.

To mitigate catastrophic forgetting during fine-tuning, examples from the pre-training dataset are replayed along the new dataset(s) \cite{Batatia2025}, and each additional dataset receives its own readout head and atomic-energy table. Datasets in practice may carry heterogeneous label sets, and hence batches are homogenized by filling missing entries with NaN sentinels, which are masked out of the loss. 

At inference time, the active head is selected by dataset name or index, optionally combined with graph-level conditions that are charge and spin multiplicity, pinning the model to the corresponding head. When using the Mixture-of-Experts formalism with eSEN, the routed expert kernels can be further contracted into a single dense kernel, recovering the runtime cost of a standard single-head model. The corresponding API is detailed in Appendix~\ref{app:api_changes}.

\subsection{Transition state search}
Locating the transition state of chemical reactions is an important task in many workflows utilizing MLIPs to model chemical reactivity. \textit{mlip} v2 includes a custom engine with an interface to the nudged elastic band (NEB) method implemented in ASE~\footnote{\url{https://ase-lib.org}}. 

In the NEB method~\cite{henkelman_2000_improved}, a set of interpolated structures, called images, between reactant and product structures of the reaction is created. Adjacent images are linked by a harmonic spring potential. The geometry of the linked images $\mathbf{R}_i$ is then optimized considering the spring forces between images $\mathbf{F}^S_{i, \parallel}$ and the actual physical forces perpendicular to the band of images $\nabla E (\mathbf{R}_{i, \perp})$,
\begin{equation}
\label{eq:neb_force}
\mathbf{F}_i = \mathbf{F}^S_{i, \parallel} - \nabla E (\mathbf{R}_{i, \perp}) \quad .
\end{equation}

After optimizing the geometry to minimize $\mathbf{F}_i$, the image with the highest potential energy can be selected as transition state guess. A useful modification is the climbing image variant of NEB \cite{henkelman_2000_a}. Here, the spring forces for the image highest in potential energy are removed, and the physical forces along the band are inverted, driving the image upwards in energy along the band to find the transition state. This variant should only be applied to a band that has already been optimized to some extent.

The NEB engine in \textit{mlip} v2 allows the user to define the number of images and the force constant between images. The NEB workflow can be started with only the structures of reactants and products, or a transition state guess can be provided as well. Prior to geometry optimization of the band, interpolation between the provided structures is done using the image dependent pair potential (IDPP) method \cite{smidstrup_2014_improved}. The NEB engine also allows providing more than three images to define an entire band, which will be optimized without interpolation. This option is especially useful for subsequent NEB simulations, for instance, using the climbing image variant. 

\subsection{NPT ensemble simulations}
\label{sec:npt}


The isothermal-isobaric (NPT) ensemble, in which the number of particles $N$, pressure $P$, and temperature $T$ are held constant, is commonly used for simulating condensed-phase systems under realistic thermodynamic conditions. In contrast to the NVT ensemble, NPT simulations allow the simulation cell to fluctuate such that the system equilibrates to a target pressure, making it particularly
important for studies of solvation, phase behavior, and biomolecular systems in explicit solvent.

\textit{mlip} v2 introduces support for NPT simulations through a JAX-based implementation of the Monte Carlo (MC) barostat~\cite{CHOW1995283, AQVIST2004288}, following the design of the OpenMM reference implementation~\cite{eastman2017}. This barostat is well-suited to MLIPs, as it performs periodic volume update proposals that are accepted or rejected using a Metropolis criterion based only on the change in potential energy and a $P\Delta V$ work term, avoiding the expensive stress evaluations required by other common barostats such as Berendsen~\cite{Berendsen1984} or Parrinello-Rahman~\cite{Parrinello1981}. When combined with the Langevin
integrator already supported for NVT simulations, this enables NPT sampling across both the ASE and JAX-MD backends. While coupling a Monte Carlo barostat with Langevin dynamics can raise questions regarding thermodynamic consistency, stochastic collisions from the Langevin heat bath and the Metropolis-Hastings criterion work together to preserve detailed balance and ensure correct statistical sampling of the phase space~\cite{liang2025efficient}. Our implementation also supports batched simulations, in which multiple systems with independent cells and random seeds evolve in parallel on a single device.
For validation of our implementation, see Appendix~\ref{app:npt}.

\section{Dataset and pre-trained models} \label{sec:results}

Along with the updated version of the library, we provide pre-trained models compatible with the updated model implementations. These models are released on HuggingFace~\footnote{\url{https://huggingface.co/collections/InstaDeepAI/ml-interatomic-potentials}} under a non-commercial license.
The models were trained on a curated version of the SPICE2 subset of OMOL25~\cite{levine2025openmolecules2025omol25} consisting of 1,763,962 structures. In Appendix~\ref{app:dataset_details}, we describe dataset cleaning, splitting strategy, and training details. The full details of the model and training hyperparameters can be found in Appendix~\ref{app:hyperparameters}.

In Figure~\ref{fig:validation_energy_forces}, we present the validation of the four pre-trained models, MACE, NequIP and VisNet, and eSEN. Each model was assessed with two standard error metrics: the mean absolute error (MAE) in predicted energies per atom (meV/atom) and the MAE in atomic forces (meV/Å). Validation was conducted across seven subsets of SPICE2, comprising 172,838 structures in total. In Appendix~\ref{app:validation}, we present additional error metrics for these models, including root-mean-squared error (RMSE) and errors for atomic partial charge prediction (MAE and RMSE). eSEN achieves the lowest energy and force errors for most subsets, while the other models exhibit very similar errors. Across most models, significantly higher energy errors are observed in the DES370K subset than in the other subset, and similarly, we obtain higher force errors in the solvated PubChem subset. As can be seen in Appendix~\ref{app:validation}, partial charge predictions are accurate across all subsets with only slightly higher errors observed for MACE and the DES370K subset. Furthermore, it should be noted that while validation errors are relevant metrics to measure training performance, they are not sufficient to attest to a model's ability to simulate correct physics~\cite{wehrhan2025mlipauditbenchmarkingtoolmachine}. Finally, we also present runtime metrics for each of the four pre-trained models in Appendix~\ref{app:validation} (see Table~\ref{tbl:performance}).

\begin{figure}[t]
    \centering
    \includegraphics[width=\linewidth]{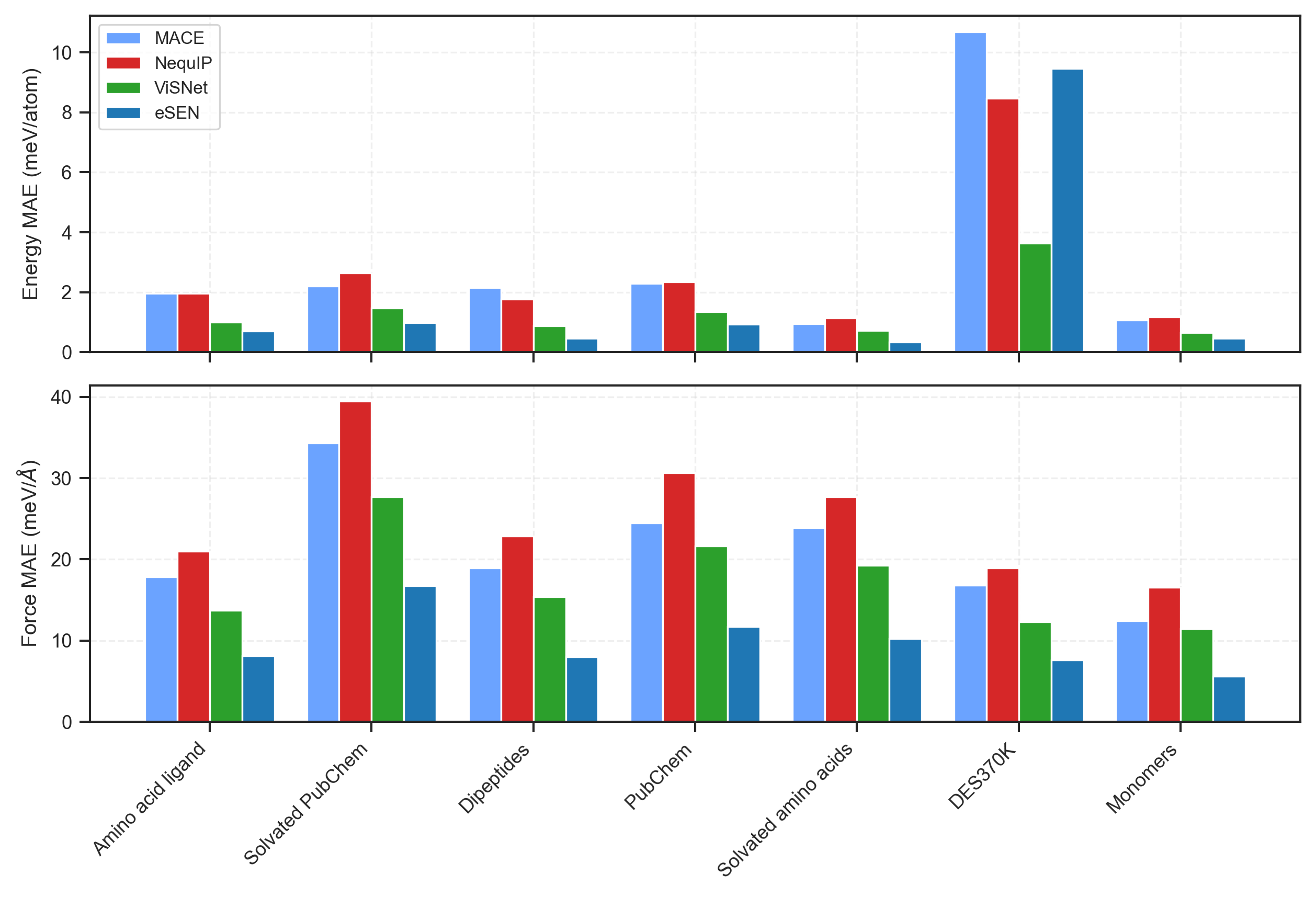}
    \caption{Validation set mean absolute errors (MAE) for energy per atom (meV/atom) and atomic forces (meV/Å) across seven molecular subsets in the SPICE2 subset of the OMOL25 dataset. The four pre-trained models MACE, NequIP, ViSNet, and eSEN are evaluated. The subsets include: amino acid ligands, solvated PubChem, dipeptides, PubChem, solvated amino acids, DES370K and monomers. MAE values reflect the deviation from DFT reference calculations. We emphasize that these results are intended to be indicative of the performance achievable with the library rather than a direct benchmark between architectures, as comparable hyperparameter settings are difficult to define.}
    \label{fig:validation_energy_forces}
\end{figure}
\section{Pre-trained models assessment by MLIPAudit}
To validate our new architectural features and the integration of data derived from the OMOL25~\cite{levine2025openmolecules2025omol25} dataset, we evaluate our pre-trained models across the standardized suite of benchmarks provided by MLIPAudit~\cite{wehrhan2025mlipauditbenchmarkingtoolmachine}. Because the primary focus of this work is to showcase the library's novel software features and implementations, this section serves as a validation check to ensure downstream consistency and physical fidelity, with comprehensive analysis deferred to the Appendix.

All four architectures, eSEN, ViSNet, MACE, and NequIP,  achieve perfect or near-perfect scores on bond length distributions, ring planarity, and reference geometry stability, and
strong scores on conformer selection (0.88--0.95) and long-timescale structural stability ($\approx$0.90), confirming physical fidelity across the model family. eSEN achieves the
highest overall score (0.716), followed by ViSNet (0.699). MACE and NequIP score similarly $\approx$0.67, in line with the obtained scores by the pre-trained models release with the v1 version of this library~\cite{brunken2025machinelearninginteratomicpotentials}. Per-benchmark scores for all models are reported in the
Appendix~\ref{app:mlipaudit}.
\section{Conclusion and outlook into future development}

In summary, \textit{mlip} v2 provides a unified, scalable, and extensible framework that significantly advances the practical applicability of MLIPs for complex molecular simulations. We are committed to maintaining the library as an open-source, actively developed project, ensuring its long-term reliability, usability, and continued alignment with advances in both machine learning and atomistic simulation. Looking ahead, we are already working on deploying additional capabilities in the near future, in particular:
\begin{itemize}
    \item Deploying MoE formalism to all implemented architectures.
    \item Including additional alternative tensor product layers such as Vector Spherical Tensor Products~\cite{luo2024enablingefficientequivariantoperations, xie2025the, xie2026asymptoticallyfastclebschgordantensor, heyraud2026integralformulasvectorspherical}.
    \item Including transformer-based MLIP architectures as some are becoming competitive in terms of runtime-accuracy trade off~\cite{liao2026equiformerv3scalingefficientexpressive, huang2026e2formerv2ontheflyequivariantattention}.
    \item Deploying functionalities for better treatment of electrostatics and long range interactions, e.g. similar to MACE-POLAR~\cite{batatia2026macepolar1polarisableelectrostaticfoundation}.
\end{itemize}

\section*{Acknowledgments}

This work was supported by Cloud TPUs from Google’s TPU Research Cloud (TRC). We would also like to thank Bohan Cao from Nankai University / Zhongguancun Academy for the numerous suggestions and conversations.

\newpage


\bibliography{references}


\newpage
\appendix

\section{Overview of API updates} \label{app:api_changes}

The following contains an overview of the most essential API updates of \textit{mlip} v2. For a more in-depth migration guide and all other details of how to use the \textit{mlip} library in general and, more specifically, the new v2 features, we refer to the code documentation.~\footnote{\url{https://instadeepai.github.io/mlip}}

\subsection*{The \texttt{ChemicalSystem} and \texttt{Graph} classes}

As mentioned in the main text, we introduce the new \texttt{Graph} class to replace the former \texttt{jraph.GraphsTuple}, which becomes the central data structure throughout all workflows. For an applied user interested in running standard training, inference, and simulation workflows, this change does not have a direct impact on their workflows. Furthermore, the already existing \texttt{ChemicalSystem} class was extended with a constructor from an \texttt{ase.Atoms} object, so that it becomes straightforward to perform inference on a single system, as demonstrated below.

\vspace*{0.2cm}
\begin{lstlisting}
import ase.io
from mlip.data import ChemicalSystem
from mlip.graph import Graph

# From molecular structure file to input graph
atoms = ase.io.read("/path/to/xyz/or/pdb/file")
chem_system = ChemicalSystem.from_ase_atoms(atoms)
graph = Graph.from_chemical_system(chem_system, graph_cutoff_angstrom=5.0)

# Generate prediction for input graph
force_field = _get_force_field_placeholder()
prediction = force_field(graph)

# Display energy prediction
print(prediction.energy)
\end{lstlisting}

Note that chemical system, graph, and prediction objects contain additional fields in v2 compared to v1 corresponding to the presented new features, e.g., partial charges or Hessian matrices. Instead of computing a \texttt{Prediction} object, the new method \texttt{ForceField.calculate} allows a user to compute an output \texttt{Graph} object directly. Internally, this output graph is transformed into a prediction via the \texttt{Graph.to\_prediction} method when the force field is called on an input graph.

\subsection*{Dataset processing}

The most significant breaking API update relates to the processing of a full dataset. In \textit{mlip} v1, the classes \texttt{ChemicalSystemReader} and \texttt{GraphDatasetBuilder} operate on three dataset splits together, one for training, validation and test set. To generalize the dataset processing code towards single-dataset inference as well as multi-dataset training and fine-tuning use cases, in v2, (i) the \texttt{ChemicalSystemReader} operates on a single split only, (ii) a new \texttt{SingleGraphDatasetBuilder} class is introduced to generate the graphs for a single split, and (iii) the \texttt{GraphDatasetBuilder} receives an updated API that allows users to easily orchestrate multiple-split use cases, not limited to the standard training use case but also including multi-dataset scenarios. Below, we provide a code example for the standard training use case.

\vspace*{0.2cm}
\begin{lstlisting}
from mlip.data import ExtxyzReader, GraphDatasetBuilder, BuilderMode

readers = {
   "train": ExtxyzReader("/path/to/train.xyz"),
   "val": ExtxyzReader("/path/to/val.xyz"), 
   "test": ExtxyzReader("/path/to/test.xyz"), 
}

builder_cfg = GraphDatasetBuilder.Config()
builder = GraphDatasetBuilder(
   readers, 
   builder_cfg, 
   BuilderMode.TRAINING, 
   dataset_info=None,  # compute from 'train' set
)

graph_datasets = builder.get_datasets()

# Access training set
train_set = graph_datasets["train"]
\end{lstlisting}

With this API, both multi-dataset training (e.g., for a Mixture-of-Experts eSEN model) and multi-head fine-tuning can be set up easily. For the former, a \texttt{readers} dictionary can be configured as follows:

\vspace*{0.2cm}
\begin{lstlisting}
readers = {
   "dataset_1": {"train": ..., "val": ...},
   "dataset_2": {"train": ..., "val": ...},
   "dataset_3": {"train": ..., "val": ...},
}
\end{lstlisting}

Dataset names can be chosen freely; the only reserved key is \textit{train}, which denotes the training split for each dataset. For multi-head fine-tuning, the setup is largely identical, with the exception that one dataset must be named \textit{replay}. This dataset represents replayed training data, from which the relevant statistics are extracted to construct the \texttt{DatasetInfo} object. Note that in both of these cases, the builder mode \texttt{BuilderMode.MULTI} should be passed to the \texttt{GraphDatasetBuilder} constructor.

\subsection*{Force field interface}

As outlined in the main text, the classes derived from \texttt{MLIPNetwork}, which implement the core model architectures, now follow a unified \texttt{Graph\,$\rightarrow$\,Graph} call signature. This design significantly increases flexibility with respect to both model inputs and outputs, allowing arbitrary features to be consumed and produced within a consistent interface. While this change is particularly relevant for model developers and researchers extending the library with new architectures or capabilities, it remains largely transparent to most applied users.

For standard usage, the primary interface continues to be the \texttt{ForceField} class, which has undergone only minor changes in its overall usage. One notable addition, however, is the introduction of the \texttt{Property} abstraction. This concept is used both to define the set of properties a model can provide and to specify the properties required by a given \texttt{ForceField}. For instance, in addition to the default predictions of energies and forces, users can request further quantities such as Hessians or partial charges in a consistent and extensible manner. The new concept is demonstrated below.

\vspace*{0.2cm}
\begin{lstlisting}
from mlip.models import Mace, ForceField
from mlip.models.model_io import load_model_from_zip
from mlip.typing.properties import Properties

# Request additional properties
# Energy and forces are requested by default
req_properties = Properties(hessian=True, partial_charges=True)

# Load pre-trained model
force_field = load_model_from_zip(
   Mace, "/path/to/pretrained_model.zip", req_properties
)

# Predict
force_field, graph = _setup_placeholder()
prediction = force_field(graph)

# For example, display partial charges
print(prediction.partial_charges)
\end{lstlisting}

For Mixture-of-Experts models (currently supported for the eSEN architecture), it is often desirable to specialize the model for inference by selecting a single expert after training, in order to improve inference efficiency. In the v2 API, this can be done straightforwardly by providing an \texttt{InferenceContext} when loading the model. This context specifies the conditions under which the model should be specialized, including information such as the training dataset, charge, and spin multiplicity.

\vspace*{0.2cm}
\begin{lstlisting}
from mlip.models import Mace, ForceField
from mlip.models.model_io import load_model_from_zip
from mlip.models.inference_context import InferenceContext

# Specialize model for neutral systems without unpaired electrons
# and, for example, for the "spice" dataset
inference_context = InferenceContext(
   charge=0,
   spin_multiplicity=1,
   dataset_name="spice",
)

# Load pre-trained model
force_field = load_model_from_zip(
   Mace, 
   "/path/to/pretrained_model.zip", 
   inference_context=inference_context,
)
\end{lstlisting}

Note that the dataset name must match the key used during multi-dataset processing, as outlined above.

\section{Training on Hessian labels} \label{app:hessian}

This section provides a brief study on the effect of training MLIP models on Hessian labels to validate our approach presented in section~\ref{sec:hessian} in the main text.

To enable a controlled and consistent evaluation of the proposed Hessian training strategy, including its implementation, inference procedure, and label processing, we constructed a dedicated dataset, including reaction data due to the relevance of accurate Hessian predictions for chemical reaction studies. The resulting dataset contains 451,000 geometries composed of four chemical elements ($\mathrm{C}$, $\mathrm{H}$, $\mathrm{N}$, and $\mathrm{O}$). It combines 116,000 systems sampled from the SPICE dataset~\cite{Eastman2023}, 126,000 from the synthetic reactive Transition1x dataset~\cite{Schreiner2022}, and 209,000 from the Reaction Graph Depth 1 (RGD1) dataset~\cite{Zhao2023}. Within these subsets, 16,000 and 26,000 Hessians were computed for systems from SPICE and Transition1x, respectively. In addition, 9,000 Hessians were calculated for reactants, products, and transition states corresponding to 3,000 reactions from the RGD1 dataset. 

The Hessian reference calculations were carried out using the \texttt{PySCF} \cite{sun2018pyscf} software at the $\omega$B97M-V/def2-\textit TZVPD level of theory.
The systems for which Hessians were evaluated span a range of molecular sizes, up to a maximum of 31 atoms. This upper bound was determined by the computational limits of the available hardware.

To ensure a fair and transparent comparison, we trained two identical MACE models using this dataset, configured as described in Table~\ref{tbl:mace_params}. The baseline model was trained using energy and force losses only, while the second model additionally incorporates supervision from the available Hessian labels. Both models were trained for 220 epochs. In the Hessian-supervised setting, eight rows were randomly sampled from each graph in a batch for Hessian loss evaluation.

For runtime assessment, we perform energy and force inference on 999 systems drawn from 333 held-out reactions, with a maximum system size of 31 atoms. The Hessian-trained model is additionally used to predict full Hessian matrices for the same set of structures. For standard energy and force inference, the average computation time per structure is $0.35$ seconds. When extending the inference to include full Hessian prediction, requiring differentiation of the force vectors with respect to all atomic positions, the average computation time increases to $0.89$ seconds on a single NVIDIA H100 GPU.

To quantitatively assess the impact of Hessian supervision, we perform a statistical comparison of vibrational frequency predictions against DFT references. Table~\ref{tbl:vib_freq_mae} reports the mean absolute errors (MAE) for a baseline MACE model and a model trained with Hessian labels. The evaluation is carried out separately for transition states and minimum-energy structures (reactants and products). The results show a clear and consistent reduction in error when Hessian information is included during training, demonstrating a significant improvement in the accuracy of MLIP-predicted vibrational frequencies.

\begin{figure}
    \centering
    \includegraphics[width=0.7\linewidth]{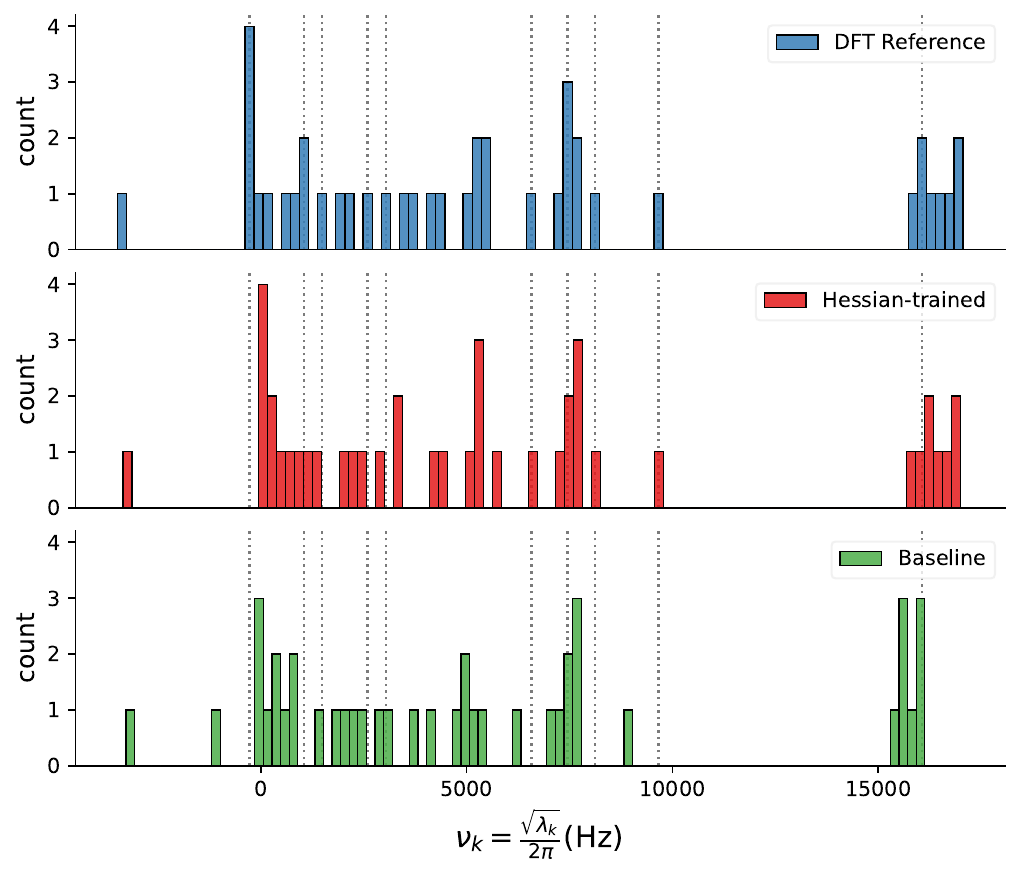}
    \caption{
        Comparison of vibrational frequencies computed with DFT, a MACE model trained on Hessian labels, and a baseline MACE model, for a single test system from the RGD1 dataset (reaction ID: \texttt{MR\_127670\_0}, chemical formula: C$_2$H$_4$N$_8$, reactant: 1-methyl-2-(prop-1-en-2-yl)diazene, product: 2-diazobutane). These results demonstrate the positive effect of the Hessian-supervised training on the alignment of the MACE predicted vibrational frequencies visually, compared to its baseline. For more statistically relevant averaged results on this dataset, see Table~\ref{tbl:vib_freq_mae}.
    }
    \label{fig:vibr_freqs_plot}
\end{figure}

To complement this aggregate analysis, we also provide a representative single-system comparison. Figure~\ref{fig:vibr_freqs_plot} illustrates the vibrational frequency spectrum for a transition state molecule from the RGD1 dataset (reaction ID: \texttt{MR\_127670\_0}, chemical formula: C$_2$H$_4$N$_8$, reactant: 1-methyl-2-(prop-1-en-2-yl)diazene, product: 2-diazobutane), showing that the spectrum predicted by the Hessian-trained MLIP is in closer agreement with the DFT reference, particularly in the high-frequency region. The vibrational frequencies $\nu_k$ are computed from the eigenvalues $\lambda_k$ of the mass-weighted Hessian matrix using the \texttt{TorchANI} package~\cite{gao2020torchani}.



\begin{table}[ht]
  \centering
  \caption{Mean absolute errors (MAE) for vibrational frequency predictions, comparing (i) a baseline MACE model and (ii) one trained with Hessian labels to the DFT reference. We distinguish between transition states and minimum-energy structures (reactants and products). We compute the shown metrics on 300 structures of each system type (hence, 600 in total) that are not in the training set. The results demonstrate that training on Hessian labels improves the accuracy of MLIP-predicted vibrational frequencies significantly.}
\begin{tabular}{lcc}
    \toprule
    \textbf{Training strategy} & \textbf{Transition States} & \textbf{Reactants and Products} \\
    \midrule
    With Hessian labels & 86.8\,Hz & 48.7\,Hz \\
    Baseline        & 253.9\,Hz & 82.9\,Hz \\
    \bottomrule
  \end{tabular}
  \label{tbl:vib_freq_mae}
\end{table}


\section{Impact of global charge embedding for training on charged systems} \label{app:global_charge}

As mentioned in the main article, we found that MLIP models have low energy accuracy for globally charged systems. While this can be partially attributed to their sparse presence in training datasets, we demonstrate in this section that using total charge information for energy prediction significantly improves MLIP model performance, particularly on globally charged systems. The global charge information is fed into the model via a learned embedding that is concatenated with the atomic number embedding and routed through a small linear MLP, as is done in the UMA architecture~\cite{wood2026umafamilyuniversalmodels}. The resulting latent vector is then used as input for the model's message passing layers.

In this section, we present a brief study illustrating the potential benefits of incorporating a global charge embedding into any model. We trained two identical ViSNet models: one with the charge embedding and one without. To ensure that globally charged systems were well represented during training, we used a subsampled version of the PMechDB dataset~\cite{Tavakolietal2024} that displayed a high occurrence of charged systems. The models were trained on approximately 120,000 structures, containing roughly 40\,\% neutral molecules, 20\,\% negatively charged molecules, and 40\,\% positively charged molecules.

\begin{figure}
    \centering
    \includegraphics[width=\linewidth]{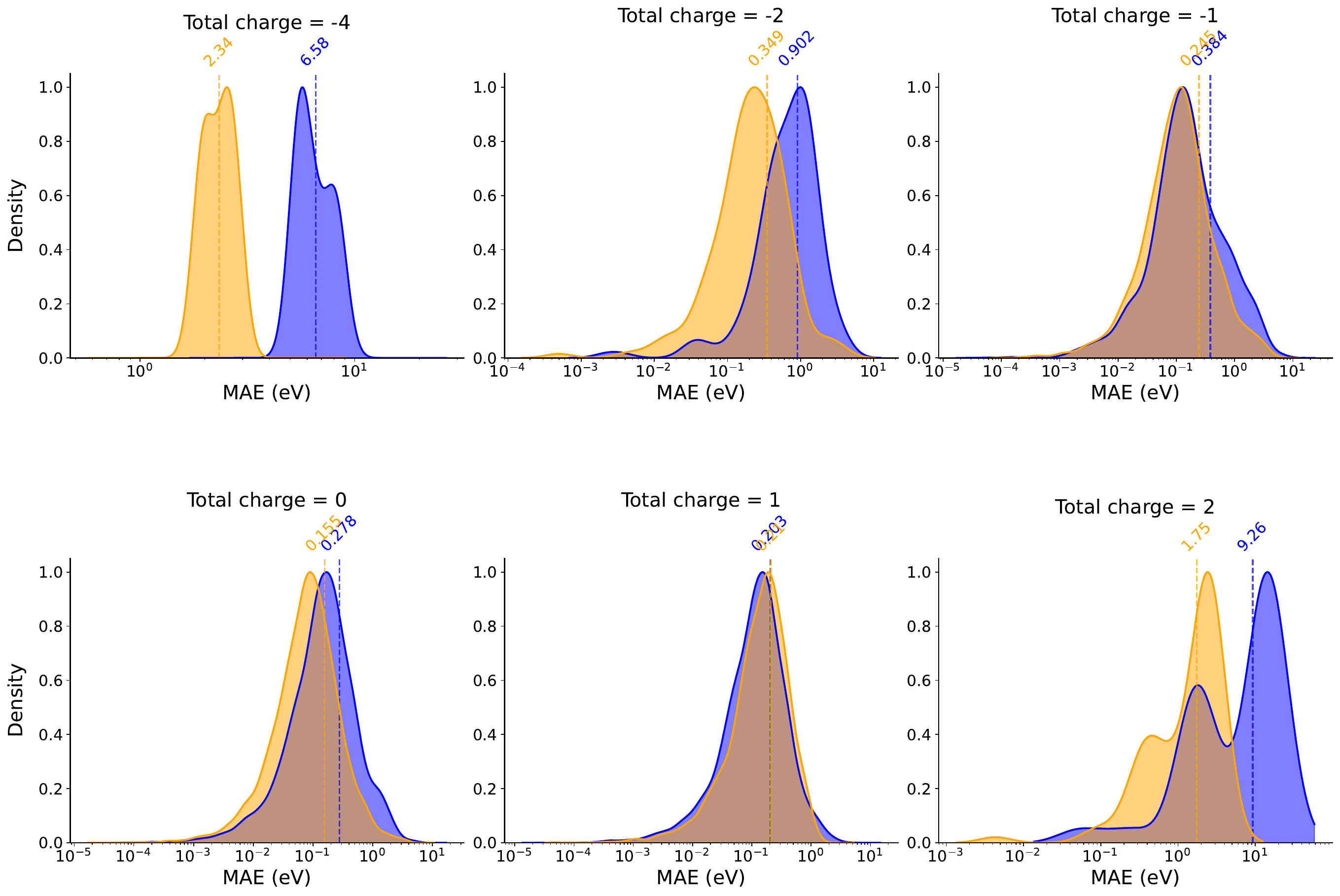}
    \caption{
        Evaluation of energy MAE distribution per global charge. A comparison of energy predictions of two equivalent ViSNet models, one trained with global charge embedding (yellow) and one without (blue), on the PmechDB dataset~\cite{Tavakolietal2024}. The distributions were obtained by evaluating the test split of the PmechDB dataset, which was used to train both models and contains around 10,000 structures with a similar charge distribution to the training split. Note that the total charge of -3 is absent from the evaluation data. The mean of each distribution is indicated by a colored vertical dashed line, and the value is displayed on top.
    }
    \label{fig:charge_embedding_plot}
\end{figure}

This experiment demonstrates that MLIP models' ability to make accurate energy predictions is affected by the global charge of the systems to some extent. As shown in Figure~\ref{fig:charge_embedding_plot}, the model's energy prediction improves significantly for highly charged systems. It is also evident that adding a global charge embedding improves the model's accuracy for neutral systems. Note that the models in Figure~\ref{fig:charge_embedding_plot} were trained on small datasets. This explains why the energy MAE of neutral systems is as high as 155\,meV, compared to the much lower numbers presented in section~\ref{sec:results} of the main text for our pre-trained models released alongside the library.

\section{Validation of JAX-based NPT integrators} \label{app:npt}
To validate the NPT integrators included in this release, which use a JAX-based Monte Carlo (MC) barostat coupled to a Langevin thermostat, we compared thermodynamic observables from the JAX-MD and ASE implementations against two reference NPT methods available in the ASE package: \textit{BerendsenNPT}, which combines a Berendsen~\cite{Berendsen1984} thermostat and barostat; and \textit{MelchionnaNPT}, which combines Nosé–Hoover thermostat with Parrinello–Rahman barostat dynamics~\cite{Melchionna1993}. All four simulations used the pre-trained VisNet model on a system of 501 water molecules at 300 K and 1 atm. The system was minimised classically using the TIP3P force field~\cite{Jorgensen1983} in OpenMM~\cite{eastman2017}, after which production runs of 500ps were performed with the last $2/3$ of each trajectory used for analysis.

\begin{figure}
    \centering
    \includegraphics[width=\linewidth]{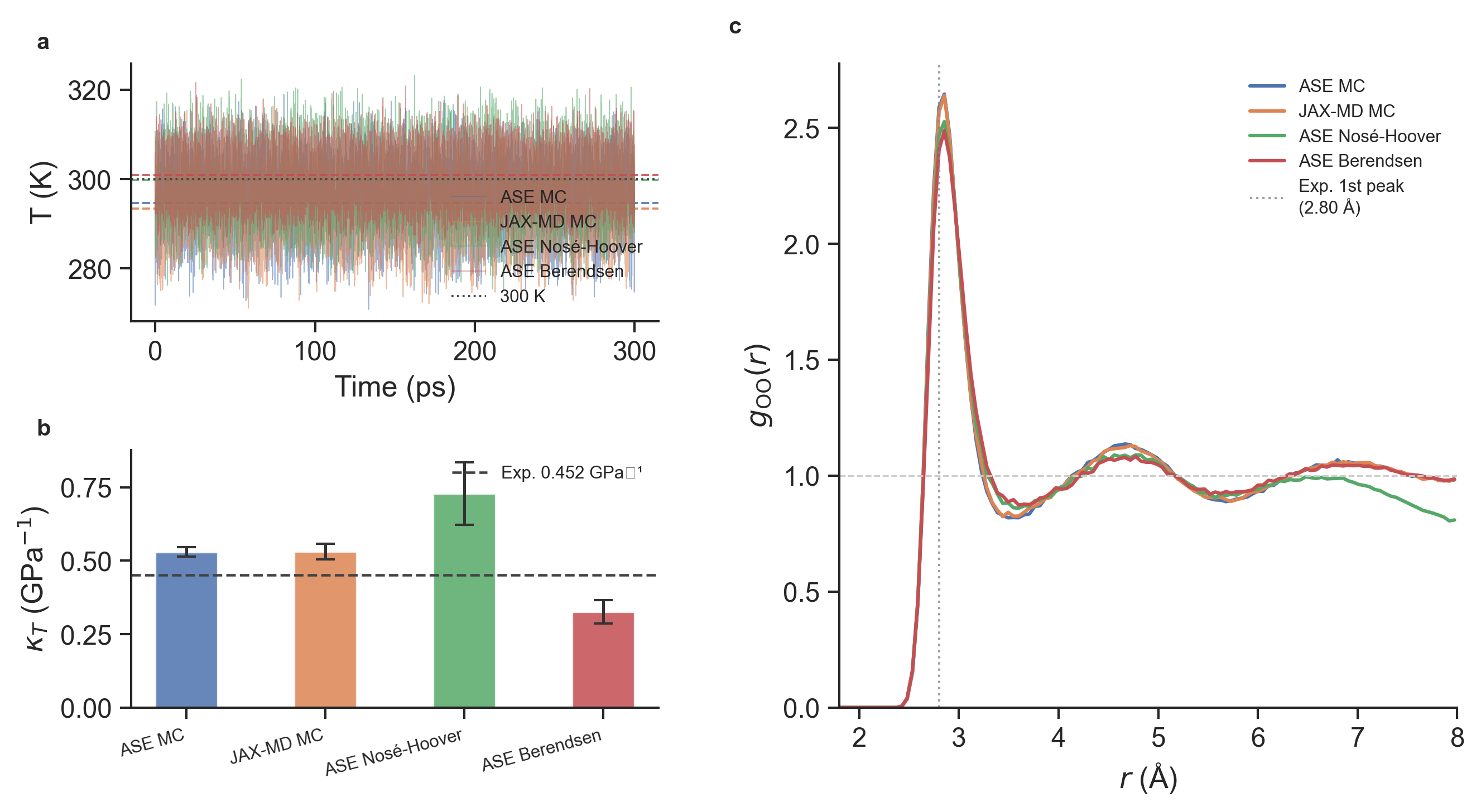}
    \caption{NPT molecular dynamics validation for VisNet-v2 water (501 molecules, 300 K, 1 atm). (a) Instantaneous temperature during the production run for each barostat; dashed lines show time averages, dotted line the target of 300 K. (b) Isothermal compressibility $\kappa_T$ estimated from volume fluctuations; error bars are standard errors from 10-block averaging; the dashed line marks the experimental value of 0.452 $GPa^{-1}$~\cite{Vega2011}.(c) Oxygen–oxygen radial distribution function $g_{OO}(r)$; the dotted vertical line indicates the experimental first-peak position at 2.80 Å~\cite{Soper2000}. 
}
    \label{fig:npt}
\end{figure}

The two JAX-based implementations show excellent agreement across all observables. Mean temperatures are consistent across all four methods (~\ref{fig:npt} panel a), each running close to the 300 K target despite differences in thermostat algorithm. The isothermal compressibility $\kappa_T$, estimated from volume fluctuations via $\kappa_T = ⟨\delta V^{2}⟩/(⟨V⟩k_{BT})$, is in quantitative agreement between the ASE and JAX-MD MC barostats (~\ref{fig:npt} panel b), with both lying above the experimental value of 0.452 $GPa^{-1}$~\cite{Vega2011}, a known overestimation attributable to the force field rather than the barostat. The Berendsen barostat yields a markedly lower $\kappa_T$, consistent with its well-documented suppression of volume fluctuations and consequent sampling of an incorrect ensemble~\cite{Berendsen1984,Bernetti2020}. The oxygen–oxygen radial distribution function $g_{OO}(r)$ is indistinguishable across all four NPT methods (~\ref{fig:npt} panel c). Together, these results establish that both JAX-based NPT integrators included in this release sample equivalent NPT statistics to alternative commonly used algorithms. Finally, we note that our JAX-MD integrator is by far the fastest evaluated in this comparison, attaining speedups of 2.2x, 2.4x and 4.0x over our ASE integrator, \textit{MelchionnaNPT} and \textit{BerendsenNPT} respectively.

\section{Dataset and training details for pre-trained models} \label{app:dataset_details}

A curated version of the SPICE2 subset of OMOL25~\cite{levine2025openmolecules2025omol25} was created by removing chemically invalid structures (based on atomic valence heuristics) and entries where either the per-atom force norm or the total force exceeded a threshold of 0.1 eV/\AA~and 15 eV/\AA, respectively. For model training, this dataset was split into training, validation and test sets using a 90:9:1 ratio. To prevent different conformers of the same molecule from being present in both sets, the split was performed on the SMILES strings. 

The final training set contains 1,763,962 structures covering 17 chemical elements ($\mathrm{B}$, $\mathrm{Br}$, $\mathrm{C}$, $\mathrm{Ca}$, $\mathrm{Cl}$, $\mathrm{F}$, $\mathrm{H}$, $\mathrm{I}$, $\mathrm{K}$, $\mathrm{Li}$,  $\mathrm{Mg}$, $\mathrm{N}$, $\mathrm{Na}$, $\mathrm{O}$, $\mathrm{P}$, $\mathrm{S}$, $\mathrm{Si}$), the validation set 172,838 structures across 17 elements, while the test set contains 17,477 structures across 13 elements. Each pre-trained model was trained for 220 epochs using NVIDIA H100 GPUs and a mean-squared-error loss. The full details of the model architectures and training hyperparameters can be found in Appendix~\ref{app:hyperparameters}.

Compared to the pre-trained models released with \textit{mlip} v1, we introduce several important modifications to both the training data and procedure. On the data side, training is now conducted on a curated version of the SPICE2 subset of OMOL25~\cite{levine2025openmolecules2025omol25}, replacing the original SPICE dataset, and now explicitly includes charged systems. The chemical coverage is also expanded, increasing the number of supported elements from 15 to 17 through the addition of $\mathrm{Ca}$ and $\mathrm{Mg}$. Furthermore, all models are extended to predict atomic partial charges alongside energies and forces.

\section{Model and training hyperparameters} \label{app:hyperparameters}
The models were trained using very similar training strategies to v1, and to each other. Training was performed over 220 epochs with scheduled MSE loss weights: energy (40) and forces (1000), flipped at epoch 115. The charge weight (1) stayed constant throughout, and the partial charge weight flipped at epoch 115 from 4 to 0.2. An exponential moving average (EMA) with decay rate 0.99 was applied. The AMSGrad variant of Adam optimizer was used. The exponential moving average of the weights is taken at every training step.  We use 4000 warmup steps followed by 360000 transition steps. Gradient clipping was performed with a norm of 500, and no gradient accumulation was applied. See Table~\ref{tbl:nequip_params} for the hyperparameters used for the NequIP model, Table~\ref{tbl:visnet_params} for ViSNet, Table~\ref{tbl:mace_params} for MACE and Table~\ref{tbl:esen_params} for eSEN. All training was done on NVIDIA H100 GPUs.
\begin{table}[ht]
  \centering
  \caption{NequIP model hyperparameters.}
  \begin{tabular}{lp{0.45\textwidth}}
    \toprule
    \textbf{Parameter} & \textbf{Value} \\
    \midrule
    \texttt{num\_layers} & \texttt{5} \\
    \texttt{target\_irreps} & \texttt{64x0e + 64x0o + 32x1o + 32x1e + 32x2e + 32x2o} \\
    \texttt{l\_max} & \texttt{2} \\
    \texttt{num\_rbf} & \texttt{8} \\
    \texttt{radial\_net\_nonlinearity} & \texttt{swish} \\
    \texttt{radial\_net\_n\_hidden} & \texttt{64} \\
    \texttt{radial\_net\_n\_layers} & \texttt{2} \\
    \texttt{radial\_envelope} & \texttt{polynomial\_envelope} \\
    \texttt{mlp\_variance\_scale} & \texttt{4.0} \\
    \texttt{embed\_activation} & \texttt{silu} \\
    \texttt{predict\_partial\_charges} & \texttt{True} \\
    \texttt{use\_total\_charge\_embedding} & \texttt{True} \\
    \texttt{use\_coulomb\_term} & \texttt{False} \\
    \texttt{graph\_cutoff\_angstrom} & 5.0 \\
    \texttt{batch\_size} & 128 \\
    \texttt{learning\_rate} & \texttt{2e-3} \\
    \bottomrule
  \end{tabular}
  \label{tbl:nequip_params}
\end{table}

\begin{table}[ht]
  \centering
  \caption{ViSNet model hyperparameters.}
  \begin{tabular}{lp{0.45\textwidth}}
    \toprule
    \textbf{Parameter} & \textbf{Value} \\
    \midrule
    \texttt{num\_layers} & \texttt{4} \\
    \texttt{num\_channels} & \texttt{128} \\
    \texttt{l\_max} & \texttt{2} \\
    \texttt{num\_heads} & \texttt{8} \\
    \texttt{num\_rbf} & \texttt{32} \\
    \texttt{trainable\_rbf} & \texttt{False} \\
    \texttt{activation} & \texttt{silu} \\
    \texttt{attn\_activation} & \texttt{silu} \\
    \texttt{embed\_activation} & \texttt{silu} \\
    \texttt{vecnorm\_type} & \texttt{None} \\
    \texttt{predict\_partial\_charges} & \texttt{True} \\
    \texttt{use\_total\_charge\_embedding} & \texttt{True} \\
    \texttt{use\_coulomb\_term} & \texttt{False} \\
    \texttt{graph\_cutoff\_angstrom} & 5.0 \\
    \texttt{batch\_size} & 128 \\
    \texttt{learning\_rate} & \texttt{1e-3} \\
    \bottomrule
  \end{tabular}
  \label{tbl:visnet_params}
\end{table}

\begin{table}[ht]
  \centering
  \caption{MACE model hyperparameters.}
  \begin{tabular}{lp{0.45\textwidth}}
    \toprule
    \textbf{Parameter} & \textbf{Value} \\
    \midrule
    \texttt{num\_layers} & \texttt{2} \\
    \texttt{num\_channels} & \texttt{128} \\
    \texttt{l\_max} & \texttt{3} \\
    \texttt{node\_symmetry} & \texttt{2} \\
    \texttt{correlation} & \texttt{2} \\
    \texttt{readout\_irreps} & \texttt{["16x0e","0e"]} \\
    \texttt{num\_readout\_heads} & \texttt{1} \\
    \texttt{num\_rbf} & \texttt{16} \\
    \texttt{activation} & \texttt{silu} \\
    \texttt{embed\_activation} & \texttt{silu} \\
    \texttt{radial\_envelope} & \texttt{polynomial\_envelope} \\
    \texttt{gate\_nodes} & \texttt{True} \\
    \texttt{include\_pseudotensors} & \texttt{False} \\
    \texttt{symmetric\_contraction\_backend} & \texttt{e3j} \\
    \texttt{predict\_partial\_charges} & \texttt{True} \\
    \texttt{use\_total\_charge\_embedding} & \texttt{True} \\
    \texttt{use\_coulomb\_term} & \texttt{False} \\
    \texttt{graph\_cutoff\_angstrom} & 5.0 \\
    \texttt{batch\_size} & 128 \\
    \texttt{learning\_rate} & \texttt{2e-3} \\
    \bottomrule
  \end{tabular}
  \label{tbl:mace_params}
\end{table}

\begin{table}[ht]
  \centering
  \caption{eSEN model hyperparameters.}
  \begin{tabular}{lp{0.45\textwidth}}
    \toprule
    \textbf{Parameter} & \textbf{Value} \\
    \midrule
    \texttt{num\_layers} & \texttt{2} \\
    \texttt{sphere\_channels} & \texttt{128} \\
    \texttt{hidden\_channels} & \texttt{128} \\
    \texttt{edge\_channels} & \texttt{128} \\
    \texttt{l\_max} & \texttt{2} \\
    \texttt{m\_max} & \texttt{2} \\
    \texttt{num\_rbf} & \texttt{32} \\
    \texttt{rbf\_type} & \texttt{gauss} \\
    \texttt{basis\_width\_scalar} & \texttt{2.0} \\
    \texttt{radial\_envelope} & \texttt{polynomial\_envelope} \\
    \texttt{trainable\_rbf} & \texttt{False} \\
    \texttt{norm\_type} & \texttt{rms\_norm\_sh} \\
    \texttt{act\_type} & \texttt{gate} \\
    \texttt{embed\_activation} & \texttt{silu} \\
    \texttt{cosine\_cutoff} & \texttt{False} \\
    \texttt{predict\_partial\_charges} & \texttt{True} \\
    \texttt{use\_total\_charge\_embedding} & \texttt{True} \\
    \texttt{use\_coulomb\_term} & \texttt{False} \\
    \texttt{graph\_cutoff\_angstrom} & 5.0 \\
    \texttt{batch\_size} & 128 \\
    \texttt{learning\_rate} & \texttt{1.5e-3} \\
    \bottomrule
  \end{tabular}
  \label{tbl:esen_params}
\end{table}

\section{MLIPAudit benchmark results}
\label{app:mlipaudit}

Table~\ref{tab:mlipaudit} reports per-benchmark scores for all four pre-trained models released with this library, following the MLIPAudit category structure: General, Small Molecules, Molecular Liquids, and Biomolecules. eSEN and ViSNet results are complete (14 scored benchmarks); MACE and NequIP results are partial at the time of writing, with sampling, stability, solvent and water radial distribution, and tautomers still in progress.

\textbf{General.} All models achieve strong stability scores ($\approx$0.90), confirming
well-behaved potential energy surfaces across the molecular systems tested.

\textbf{Small Molecules.} This is the most densely evaluated category. All models score
perfectly on bond length distributions and ring planarity, and achieve near-perfect scores
on reference geometry stability (1.000 for eSEN, MACE, and NequIP; 0.995 for ViSNet).
Conformer selection scores range from 0.875 to 0.946, and dihedral scan scores from 0.633
to 0.704. Non-covalent interactions are moderate (0.548--0.652), with eSEN leading. The
most challenging benchmarks in this category are tautomers (0.07--0.10), nudged elastic
band (0.16--0.24), and reactivity (0.21--0.28), all of which require accurate treatment of
transition states and proton-transfer processes that are underrepresented in current
training data.

\textbf{Molecular Liquids.} eSEN and ViSNet both reproduce water and solvent radial
distribution functions with high fidelity (0.936--1.000), indicating well-calibrated
short-range interactions in condensed-phase environments.

\textbf{Biomolecules.} Folding stability scores cluster around 0.53 across all models,
suggesting systematic underestimation of long-range conformational preferences. Sampling
scores of 0.744 (eSEN) and 0.781 (ViSNet) indicate physically meaningful free-energy
landscapes for small peptides.

\begin{table}[h]
\centering
\caption{MLIPAudit benchmark scores (fast run mode), grouped by category. $-$ denotes
benchmarks not yet completed; n/s denotes benchmarks without a numeric score.}
\label{tab:mlipaudit}
\begin{tabular}{lcccc}
\toprule
Benchmark & ESEN & ViSNet & MACE$^\dagger$ & NequIP$^\dagger$ \\
\midrule
\multicolumn{5}{l}{\textit{General}} \\
Stability                      & 0.900 & 0.900 &   $-$ &   $-$ \\
Scaling                        &  n/s  &  n/s  &  n/s  &  n/s  \\
\midrule
\multicolumn{5}{l}{\textit{Small Molecules}} \\
Bond length distribution       & 1.000 & 1.000 & 1.000 & 1.000 \\
Ring planarity                 & 1.000 & 1.000 & 1.000 & 1.000 \\
Reference geometry stability   & 1.000 & 0.995 & 1.000 & 1.000 \\
Conformer selection            & 0.946 & 0.913 & 0.898 & 0.875 \\
Dihedral scan                  & 0.704 & 0.687 & 0.640 & 0.633 \\
Noncovalent interactions       & 0.652 & 0.566 & 0.561 & 0.548 \\
Nudged elastic band            & 0.200 & 0.180 & 0.160 & 0.240 \\
Reactivity                     & 0.268 & 0.217 & 0.278 & 0.211 \\
Tautomers                      & 0.099 & 0.074 &   $-$ &   $-$ \\
\midrule
\multicolumn{5}{l}{\textit{Molecular Liquids}} \\
Water radial distribution      & 0.995 & 1.000 &   $-$ &   $-$ \\
Solvent radial distribution    & 0.995 & 0.936 &   $-$ &   $-$ \\
\midrule
\multicolumn{5}{l}{\textit{Biomolecules}} \\
Folding stability              & 0.529 & 0.532 & 0.531 & 0.529 \\
Sampling                       & 0.744 & 0.781 &   $-$ &   $-$ \\
\midrule
\textbf{Overall}               & \textbf{0.716} & \textbf{0.699} & 0.674 & 0.671 \\
\bottomrule
\end{tabular}
\end{table}

$^\dagger$ Partial results: overall scores averaged over completed benchmarks only.

\section{Additional validation and performance metrics for pre-trained models} \label{app:validation}

\begin{figure}[ht]
    \centering
    \includegraphics[width=\linewidth]{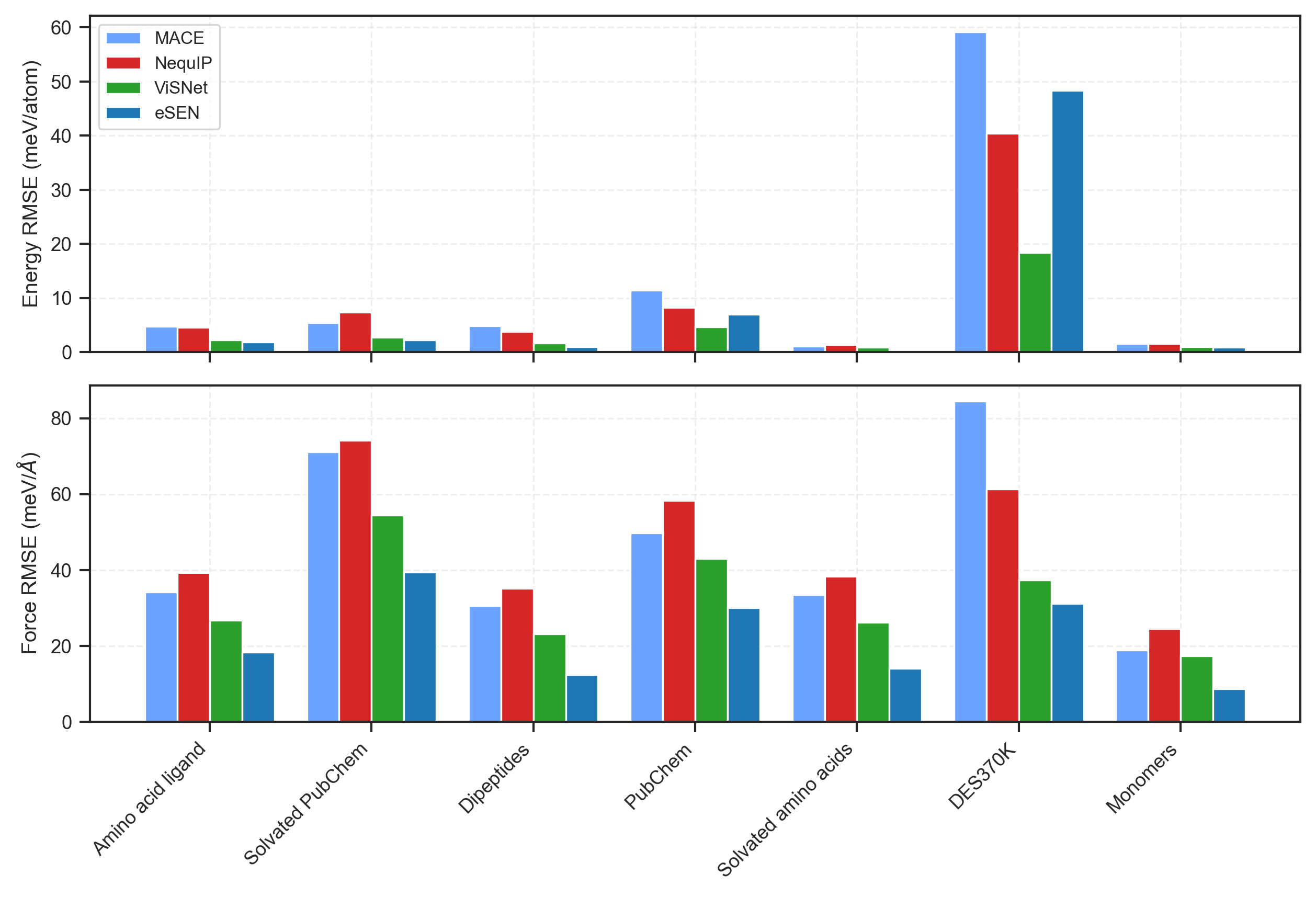}
    \caption{Validation set root-mean-squared errors (RMSE) for energy per atom (meV/atom) and atomic forces (meV/Å) across seven molecular subsets in the SPICE2 subset of the OMOL25 dataset. The four pre-trained models MACE, NequIP, ViSNet, and eSEN are evaluated. The subsets include: amino acid ligands, solvated PubChem, dipeptides, PubChem, solvated amino acids, DES370K and monomers. RMSE values reflect the deviation from DFT reference calculations.}
    \label{fig:validation_rmse}
\end{figure}

\begin{figure}[ht]
    \centering
    \includegraphics[width=\linewidth]{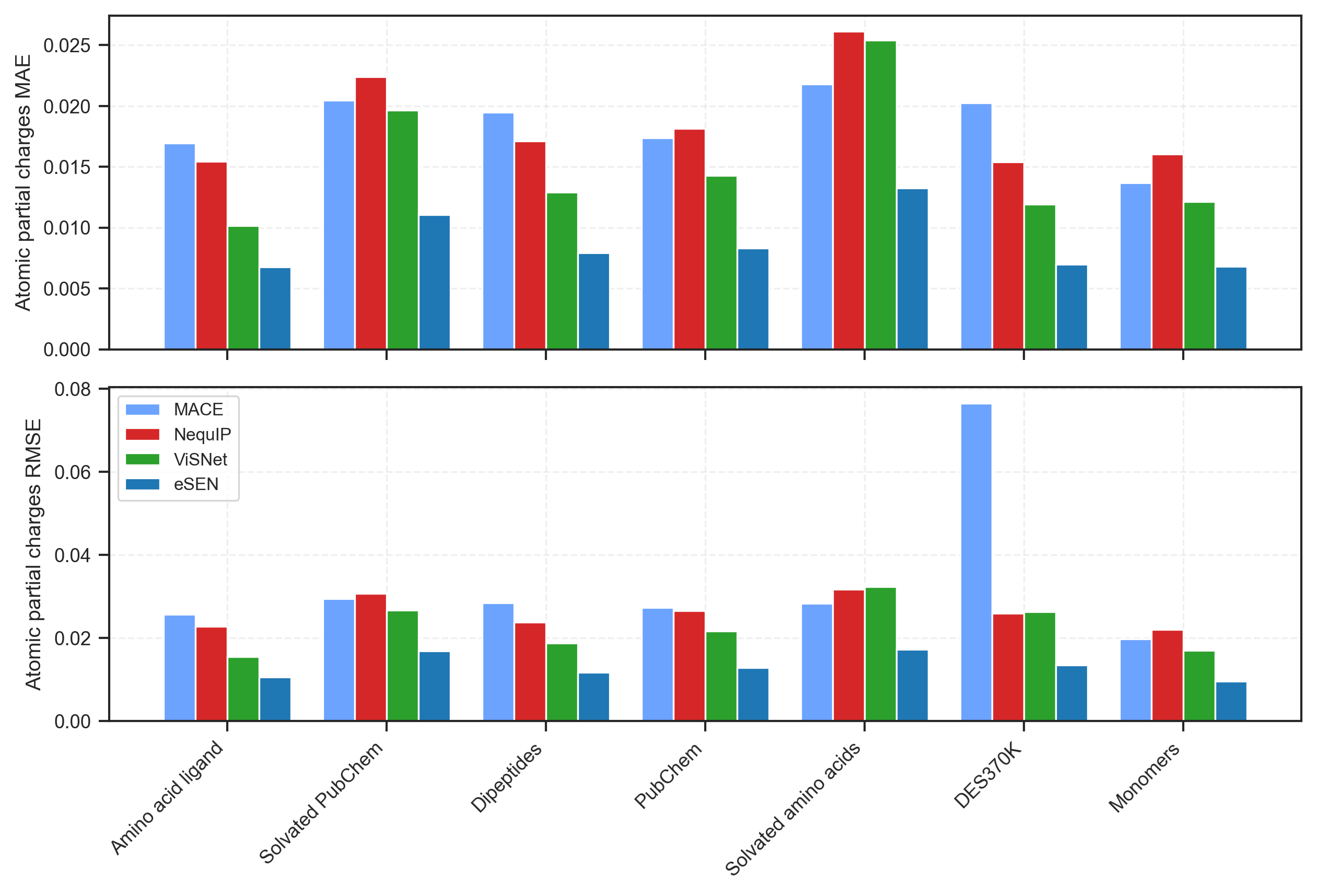}
    \caption{Validation set mean absolute errors (MAE) and root-mean-squared errors (RMSE) for atomic partial charge predictions across seven molecular subsets in the SPICE2 subset of the OMOL25 dataset. The four pre-trained models MACE, NequIP, ViSNet, and eSEN are evaluated. The subsets include: amino acid ligands, solvated PubChem, dipeptides, PubChem, solvated amino acids, DES370K and monomers. MAE and RMSE values reflect the deviation from DFT reference calculations.}
    \label{fig:validation_partial_charges}
\end{figure}

In Figure~\ref{fig:validation_rmse}, we present the corresponding RMSE error metrics for the MAE metrics presented in Figure~\ref{fig:validation_energy_forces} of the main text. Furthermore, Figure~\ref{fig:validation_partial_charges} shows the MAE and RMSE metrics for the partial charge predictions of the pre-trained models (MACE, NequIP, ViSNet, and eSEN).

Furthermore, we evaluate the runtime performance of the different pre-trained models on MD simulations using both the JAX-MD and ASE backends (Table~\ref{tbl:performance}). All benchmarks were conducted on a single NVIDIA H100 GPU, and the results span two representative systems: 1UAO, a small chignolin molecule comprising 138 atoms, and 1ABT with 1205 atoms. More details about these test systems, as well as their visualization can be found in the \textit{mlip} v1 publication~\cite{brunken2025machinelearninginteratomicpotentials}. Across both backends and system sizes, all evaluated models maintain stable MD trajectories. We note that the reported runtimes are not directly comparable to their \textit{mlip} v1 counterparts due to differences in hyperparameter settings.

\begin{table}[ht]
  \centering
  \caption{Speed performance on MD simulation of the different pre-trained models for the JAX-MD and ASE simulation backends. All tests were run on a single NVIDIA H100 GPU, and speed metrics are given in milliseconds per step, averaged over 1\,ns of simulation. 1UAO is a chignolin molecule with 138 atoms, while 1ABT is a system with 1205 atoms. All models included in the table achieved stable simulations on these benchmarks. Note that the model runtimes cannot be compared directly to their \textit{mlip} v1 counterparts due to different hyperparameters.}
  \begin{tabular}{ll|crr}
    \toprule
       Models             & Parameters & Systems & JAX-MD & ASE\\
  \midrule
  \multirow{2}{*}{MACE} & \multirow{2}{*}{3,274,016} & 1UAO   & 2.4 ms/step & 7.3 ms/step\\
                        &                            & 1ABT   & 19.2 ms/step & 43.8 ms/step\\
  \midrule
  \multirow{2}{*}{NequIP} & \multirow{2}{*}{1,921,280} & 1UAO   & 3.4 ms/step & 8.9 ms/step\\
                          &                            & 1ABT   & 22.0 ms/step & 44.6 ms/step\\
  \midrule
  \multirow{2}{*}{ViSNet} & \multirow{2}{*}{1,172,676} & 1UAO   & 1.9 ms/step & 7.1 ms/step\\
                          &                            & 1ABT   & 13.7 ms/step & 30.2 ms/step\\
  \midrule
  \multirow{2}{*}{eSEN} & \multirow{2}{*}{3,210,498} & 1UAO   & 3.0 ms/step & 8.9 ms/step \\
                        &                            & 1ABT   & 22.8 ms/step & 46.7 ms/step\\
  \bottomrule
  \end{tabular}
  \label{tbl:performance}
\end{table}


\clearpage
\newpage

\end{document}